\documentclass[a4paper,11pt]{article}
\usepackage{pos}
\usepackage{amsmath}
\usepackage{tabularx}
\def\beq{\begin{equation}}
\def\eeq{\end{equation}}
\def\bea{\begin{eqnarray}}
\def\eea{\end{eqnarray}}
\usepackage{slashed}

\usepackage{natbib}
\makeatletter
\g@addto@macro\bfseries{\boldmath}
\makeatother

\usepackage{
amstext,amsbsy,amsmath,amsfonts}
\usepackage{nicefrac}
\usepackage{MnSymbol}
\usepackage{enumitem}

\DeclareMathOperator{\im}{Im}

\def\beq{\begin{equation}}
\def\eeq{\end{equation}}
\def\bea{\begin{eqnarray}}
\def\eea{\end{eqnarray}}
\def\beqa{\begin{equation}\begin{array}{l}}
\def\eeqa{\end{array}\end{equation}}
\def\eqlab#1{\label{eq:#1}}

\def\seclab#1{\label{sec:#1}}

\def\Eqref#1{Eq.~(\ref{eq:#1})}

\def\Figref#1{Fig.~\ref{fig:#1}}

\def\secref#1{Sec.~\ref{sec:#1}}

\def\barr{\left(\begin{array}{c}}
\def\earr{\end{array}\right)}
\def\bmat{\left(\begin{array}{cc}}
\def\emat{\end{array}\right)}
\def\al{\alpha}

\def\ga{\gamma} 
\def\de{\delta}

\def\dd{\mathrm{d}}

\def\ol#1{\overline{#1}}

\def\2PE{2$\upgamma$}
\def\mTPE{{2\upgamma}}

\usepackage{upgreek}

\title{Theoretical discrepancies in the nucleon spin structure and the hyperfine splitting of muonic hydrogen}
\ShortTitle{Theoretical discrepancies in the nucleon spin structure and $\mu$H hfs}

\author*[a]{Vladimir Pascalutsa}
\author[a,b]{Franziska Hagelstein}
\author[a]{Vadim Lensky}

\affiliation[a]{Johannes-Gutenberg Universit\"at Mainz, D-55099 Mainz, Germany}
\affiliation[b]{Paul Scherrer Institut, CH-5232 Villigen PSI, Switzerland}
\emailAdd{hagelste@uni-mainz.de}
\emailAdd{vlenskiy@uni-mainz.de}
\emailAdd{pascalut@uni-mainz.de}

\abstract{Two groups, ours (Mainz) and Bochum, have recently been  re-evaluating the  
spin polarizabilities and spin structure functions at low $Q$, using the
baryon chiral perturbation theory (B$\chi$PT), the manifestly-covariant counterpart of the heavy-baryon chiral perturbation theory (HB$\chi$PT). 
Whilst the two groups agree that the B$\chi$PT framework works better than HB$\chi$PT in this sector, their quantitative results disagree 
in some of the quantities; most notably, the proton spin polarizabilities $\gamma_0$ and $\delta_{LT}$. These discrepancies are
especially intriguing in light of new experimental data coming from the Jefferson Lab ``Spin Physics Program''. The preliminary data
on the proton are reported 
by Karl Slifer in a plenary session of this workshop. 

Another theoretical discrepancy is emerging in the proton-polarizability 
contribution to the hyperfine splitting (hfs) in hydrogen and muonic hydrogen. Our B$\chi$PT calculation shows a significantly smaller effect than
the state-of-the-art data-driven evaluations based on empirical spin structure functions. The smaller polarizability contribution leads to a smaller
Zemach radius of the proton. This discrepancy could be relevant for the planned first-ever measurement of the ground-state hfs in muonic hydrogen. }

\FullConference{%
  The 10th International Workshop on Chiral Dynamics - CD2021\\
  15-19 November 2021\\
  Online
  }

\begin{document}
\maketitle

\begin{figure}[b]
\centering
  \includegraphics[trim=0cm 0.9cm 11cm 0cm,clip,scale=0.5]{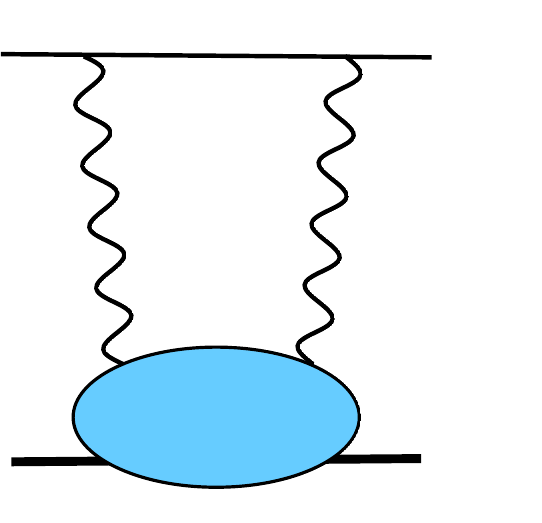} \hskip3cm
  \includegraphics[scale=0.5]{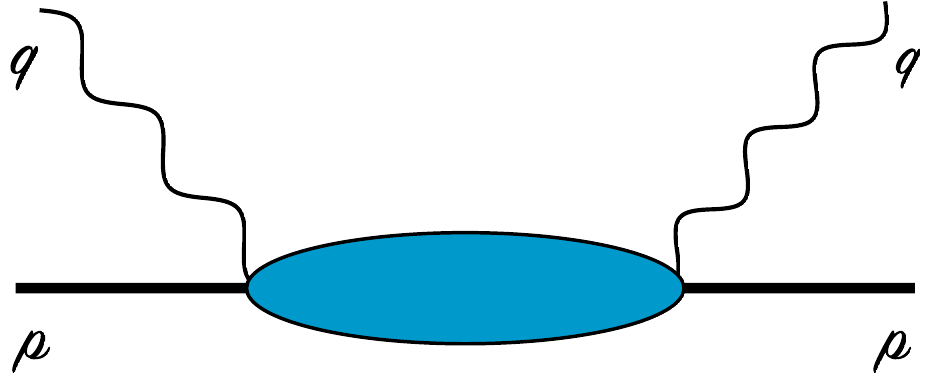}
\caption{Left: \2PE-exchange diagram; horizontal lines correspond to the lepton and the nucleus (lower, bold). Right: forward Compton scattering.   \label{fig:TPE}} \label{fig:VVCS}
\end{figure}

\section{Introduction}

Compton scattering off nucleons ($\gamma N \to \gamma N$) is one of the simplest processes 
by which we access the electromagnetic 
structure of the nucleon. It plays a 
central role in the calculation of the two-photon-exchange (2$\gamma$-exchange) corrections, Fig.~\ref{fig:TPE}, in hydrogen (H) and muonic hydrogen ($\mu$H), 
as well as, of the radiative corrections to elastic lepton-nucleon
scattering (see, e.g., \cite{Hagelstein:2015egb,Pasquini:2018wbl} for reviews).

The description of nucleon structure at low energy relies on data-driven dispersive approaches, lattice QCD, and low-energy effective field theories (EFTs). Here we focus on the latter. We employ the baryon chiral perturbation theory (B$\chi$PT) \cite{Gasser:1987rb,Fuchs:2003qc,Pascalutsa:2003aa} --- an EFT operating in terms of pion, nucleon and $\Delta$-isobar fields --- to compute the forward doubly-virtual Compton scattering (VVCS) on the nucleon at next-to-next-to-leading order \cite{Lensky:2014dda,Alarcon:2020wjg,Alarcon:2020icz}. An analogous calculation is done by the Bochum group \cite{Bernard:2012hb,Thurmann:2020mog}, with insofar inexplicable differences to our results 
(compare, e.g., blue and grey bands in \Figref{protonspin}).

The VVCS amplitudes contain the main ingredients of the nucleon electromagnetic structure: form factors, polarizabilities, structure functions, which is useful in evaluations of the aforementioned
2$\gamma$-exchange corrections.
For example, the nucleon spin structure functions
$g_1$ and $g_2$ are important for extractions of nuclear Zemach radii from the hyperfine-splitting (hfs) measurements in light muonic atoms.
 Presently, several collaborations (CREMA~\cite{Amaro:2021goz}, FAMU~\cite{Pizzolotto:2021dai, Pizzolotto:2020fue} and J-PARC/Riken~\cite{Sato:2014uza}) are preparing measurements of the ground-state hfs in $\mu$H and $\mu^3$He$^+$. These will allow one to extract the Zemach radii of the proton and helion, and learn about their magnetic properties. For a successful measurement, precise theory predictions are needed to narrow down the frequency search range in the experiments, see~\cite{Antognini:2022xoo} for review. In light muonic atoms, these predictions  are usually limited by the uncertainty of the nuclear- and nucleon-structure effects, which mainly stem from  the \2PE-exchange to be discussed in this paper. 

The paper  is organized as follows. In \secref{Pol}, we show the B$\chi$PT predictions for the nucleon spin polarizabilities and moments of polarized structure functions and compare them to experimental data which have been recently obtained at the Jefferson Lab. In \secref{hfsChPTSEC}, we discuss the leading-order B$\chi$PT predictions of the \2PE-exchange polarizability contributions to the hfs in H and $\mu$H, and compare them to data-driven dispersive evaluations. We conclude with a summary and conclusions in \secref{Sum}.

\section{Nucleon Spin Polarizabilities and Moments of Polarized Structure Functions} \seclab{Pol}

In the two recent  papers~\cite{Alarcon:2020wjg,Alarcon:2020icz}, we discussed both the spin-independent and spin-dependent VVCS amplitudes, $ T_{1,2}(\nu,Q^2)$ and $S_{1,2}(\nu,Q^2)$, functions of the photon energy $\nu$ and virtuality $Q^2$. The absorptive part of these amplitudes, via the optical theorem, yields the  unpolarized and polarized  structure functions, $\im T_i (\nu,Q^2) \sim F_i(x,Q^2)$ and $\im S_i (\nu,Q^2) \sim g_i(x,Q^2)$, with $x=Q^2/(2M_N \nu)$ the Bjorken variable and  $M_N$ the nucleon mass. 
Here we will focus on the polarized observables,
some of which have recently received new experimental data. 

The nucleon spin structure functions are being measured at Jefferson Lab  within the ``Spin Physics Program'' \cite{Chen:2008ng}. Some new results have been presented in this workshop (see, e.g.~\cite{Deur:2022sot} and the plenary contribution of K.~Slifer). At least three different  experiments have recently been mapping out the spin structure functions of the nucleon over a wide kinematic range: the EG4 experiments by the CLAS Collaboration (E03-006 for the proton and E06-017 for the neutron using NH$_3$ and ND$_3$ targets) \cite{CLAS:2017ozc,CLAS:2021apd}, the E97-110 experiment (using a $^3$He target to study the neutron) \cite{JeffersonLabE97-110:2019fsc,E97-110:2021mxm}, and  the E08-027 or g2p experiment (using an NH$_3$ target to study the proton) \cite{Zielinski:2017gwp,JeffersonLabHallAg2p:2022qap}. 

\begin{figure}[bth]
\begin{center}
             \centering     \includegraphics[height=5.1cm]{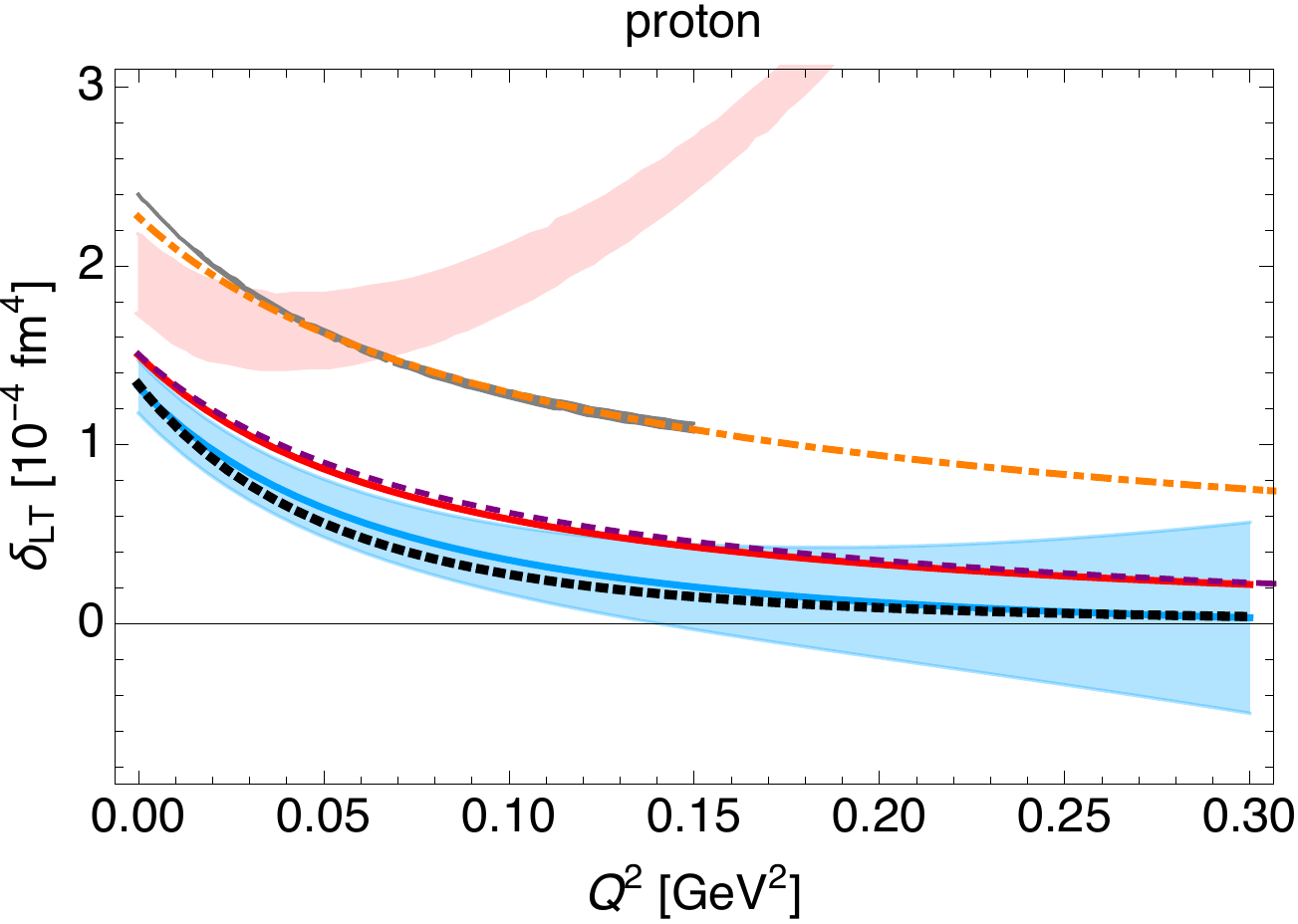}
      \hfill
            \includegraphics[height=5.1cm]{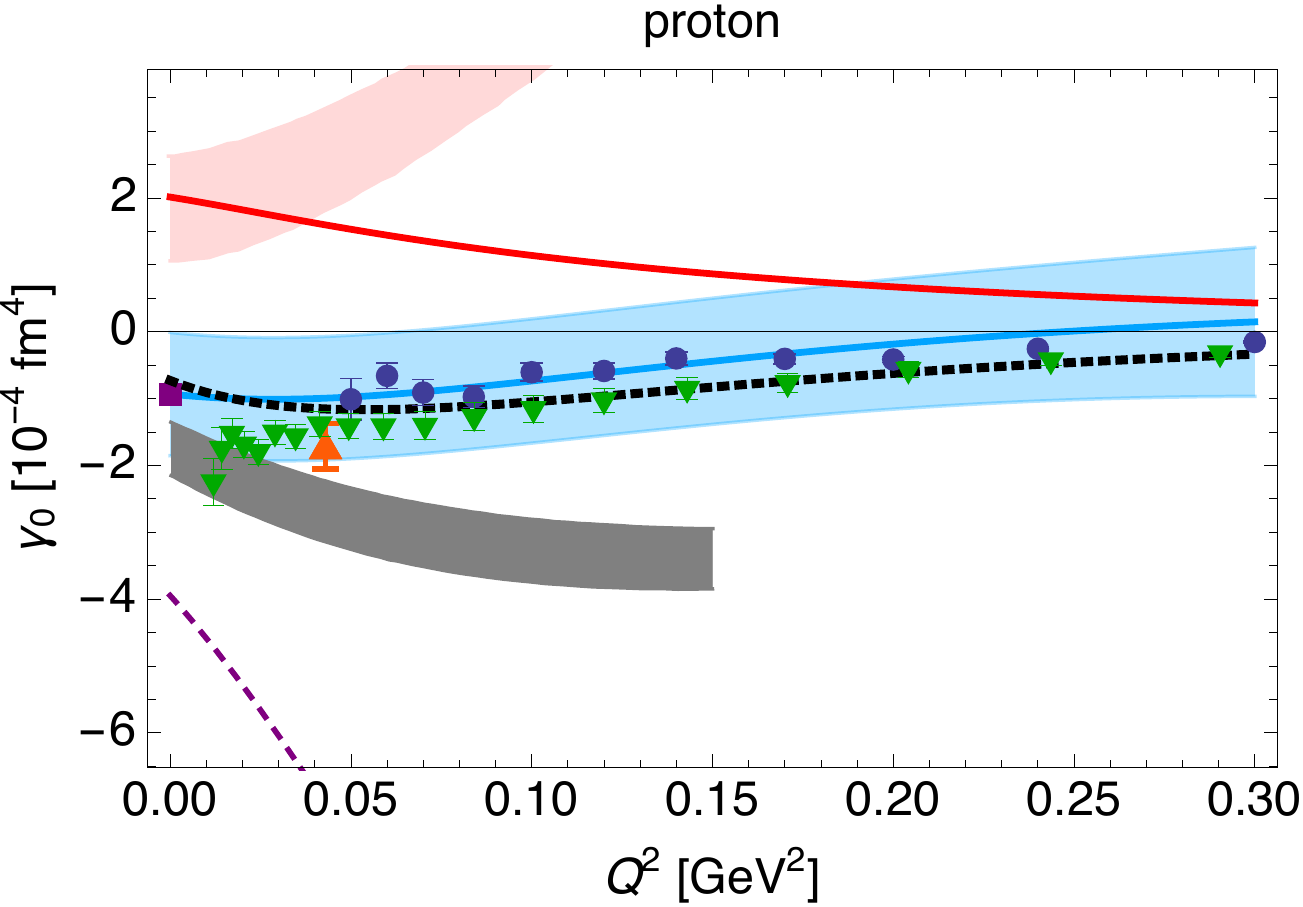}
            \vskip2mm
             \includegraphics[height=5.1cm]{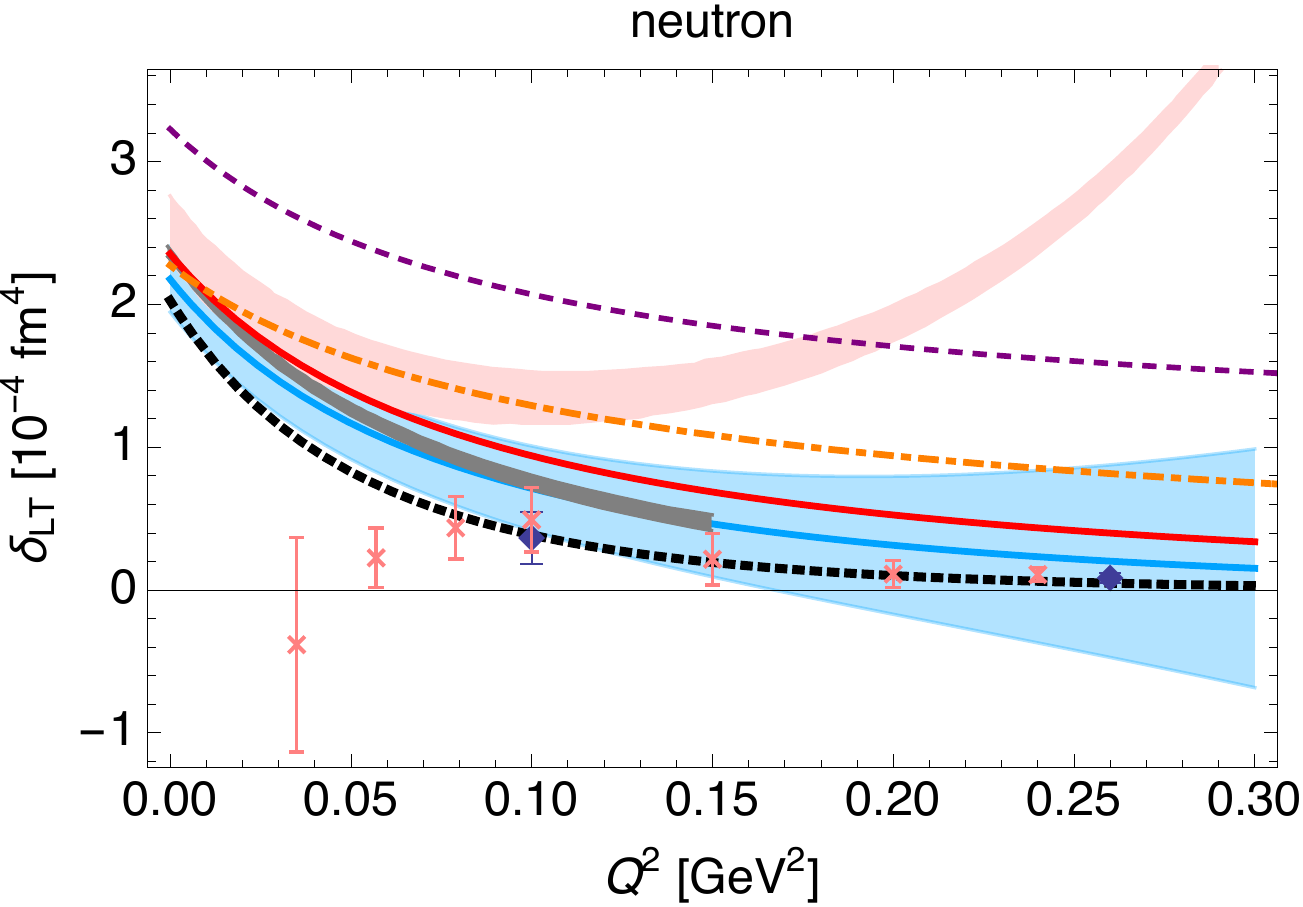}
    \hfill
\includegraphics[height=5.1cm]{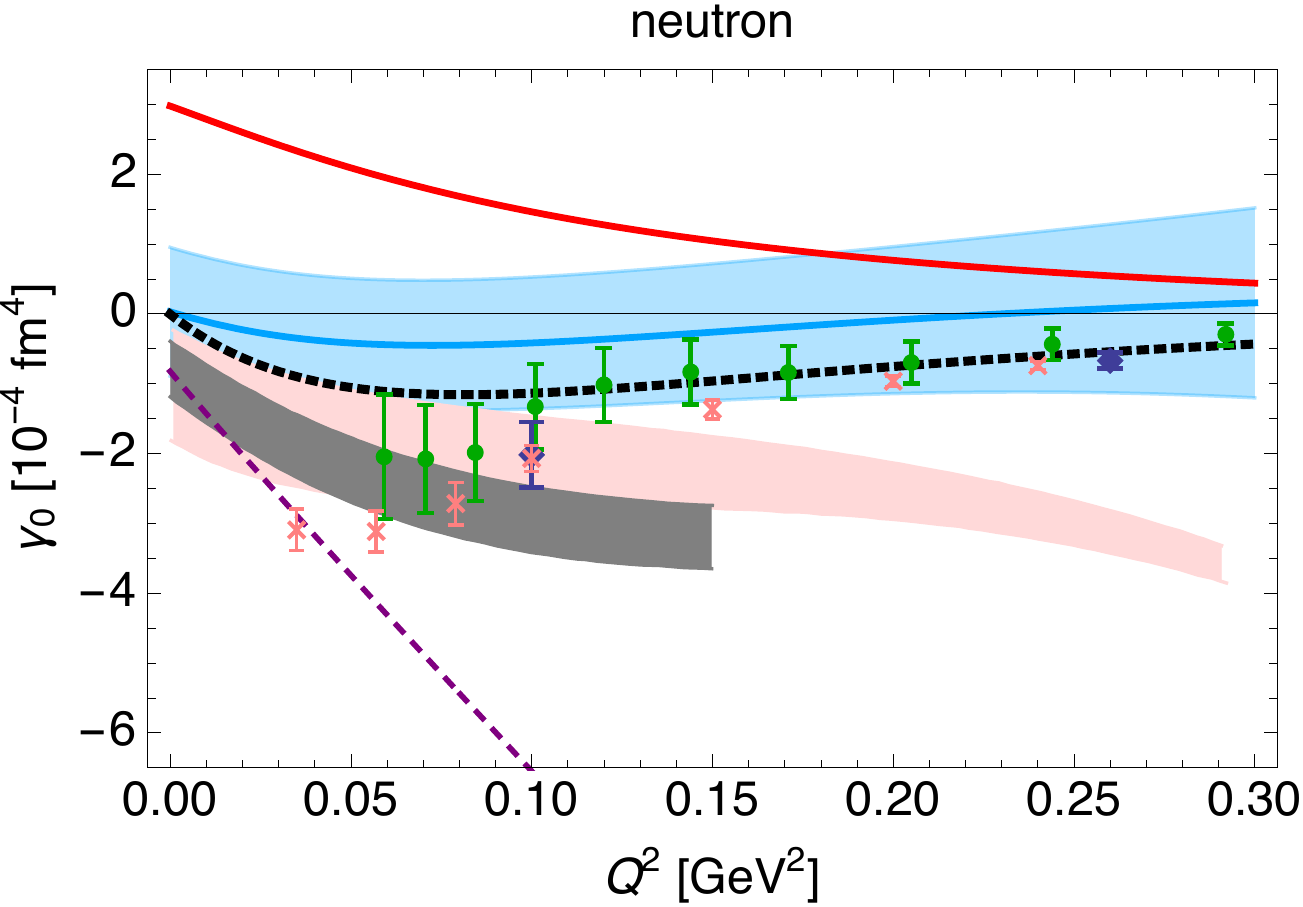}
\caption{Left: longitudinal-transverse spin polarizability $\delta_{LT}(Q^2)$. Right: forward spin polarizability $\gamma_0(Q^2)$. Upper: proton. Lower: neutron. 
The red curves represent the $\mathcal{O}(p^3)$ B$\chi$PT. The orange dot-dashed and purple dashed curves represent, respectively,  $\mathcal{O}(p^3)$ and $\mathcal{O}(p^4)$ HB$\chi$PT~\cite{Kao:2002cp,Kao:2003jd}.
 The pink bands represent the IR-$\chi$PT~\cite{Bernard:2002pw}. The gray bands represent the Bochum B$\chi$PT~\cite{Bernard:2012hb}. Our B$\chi$PT results \cite{Alarcon:2020icz} are shown by the blue band.
 The black-dotted curves represent the MAID model with $\pi,\eta,\pi\pi$ channels \cite{Drechsel:2002ar}.
Data-driven evaluations: \cite{Prok:2008ev} blue dots, \cite{Gryniuk:2016gnm} purple square for $\gamma_0(0)$, \cite{Zielinski:2017gwp} orange pyramid, \cite{CLAS:2021apd} green triangles, \cite{Amarian:2004yf} blue diamonds, \cite{CLAS:2015otq} green dots,  and \cite{E97-110:2021mxm} pink crosses. 
 \label{fig:protonspin}   }
\end{center}
\end{figure}

We consider, first of all,  the following spin
polarizabilities:
\begin{subequations}
\eqlab{SRintegrals}
\bea
\gamma_0 (Q^2)
&=&\frac{16 \al M_N^2}{Q^6}\int_0^{x_0}\!\dd x \, x^2 \! \left[g_1(x,Q^2)-\frac{4M_N^2 x^2}{Q^2}\,g_2(x,Q^2)\right]\!,\eqlab{Eq:gamma0Q2}\\
\delta_{LT} (Q^2)
&=&\frac{16\al M_N^2}{Q^6}\!\int_0^{x_0}\!\dd x \, x^2\left[g_1(x,Q^2)+g_2(x,Q^2)\right]\!,\quad\eqlab{Eq:deltaLTQ2}
\eea
\end{subequations}
where $\al\simeq 1/137.036$ is the fine structure constant, and $x_0$ is the inelastic threshold. They are shown in \Figref{protonspin} for the proton (upper panels) and neutron (lower panels). 
The red curves show the leading, $O(p^3)$ prediction of B$\chi$PT (see the Feynman diagrams in Fig.~1 of Ref.~\cite{Lensky:2014dda}). The blue curves and error bands show the next-to-leading,
 $O(p^{7/2})$ prediction, which includes in addition the $\Delta(1232)$-isobar contributions (Fig.~2 of Ref.~\cite{Lensky:2014dda}). The gray bands show the  analogous B$\chi$PT calculation by the Bochum group~\cite{Bernard:2012hb}. Apparently there is a significant discrepancy between the two B$\chi$PT calculations, especially in the proton $\de_{LT}$.

 Other theory predictions seen in the figure include the considerably older HB$\chi$PT calculations. They demonstrate the poor convergence
 of HB$\chi$PT for these quantities (some of the curves are outside the scale of the figure). The pink bands represent the calculation in ``infrared regularization" scheme of B$\chi$PT, exhibiting unphysical singularities which make it even less viable than HB$\chi$PT.

Coming back to the discrepancy between the two B$\chi$PT calculations, let us note that one of them (blue bands) comes out to be fairly consistent with the empirical evaluation of MAID \cite{Drechsel:2002ar}, represented by the dotted curves in the figure.
 MAID uses cross sections for individual channels, rather than the total inclusive. However, the integral over $x$ for these quantities converges very rapidly for low $x$ (high energies), which makes the contribution of other channels negligible. Conversely, any large deviation from MAID, as seen, for example, by the Bochum calculation (grey bands)
 for the proton, 
 should be reproduced by large high-energy contributions on the side of data-driven evaluations. 
 
 \begin{figure}[th]
\begin{center}
 \includegraphics[height=5cm]{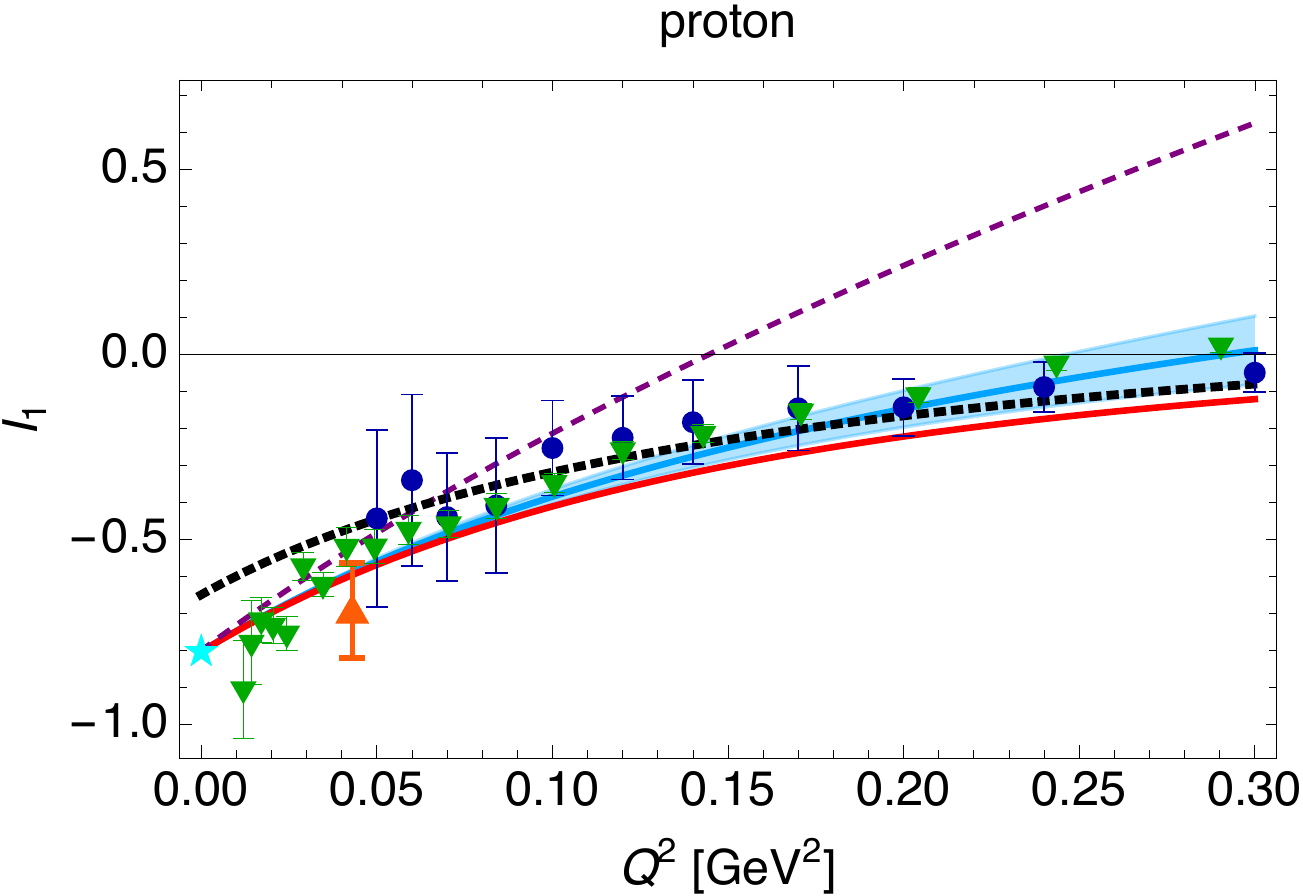}
    \hfill
     \includegraphics[height=5cm]{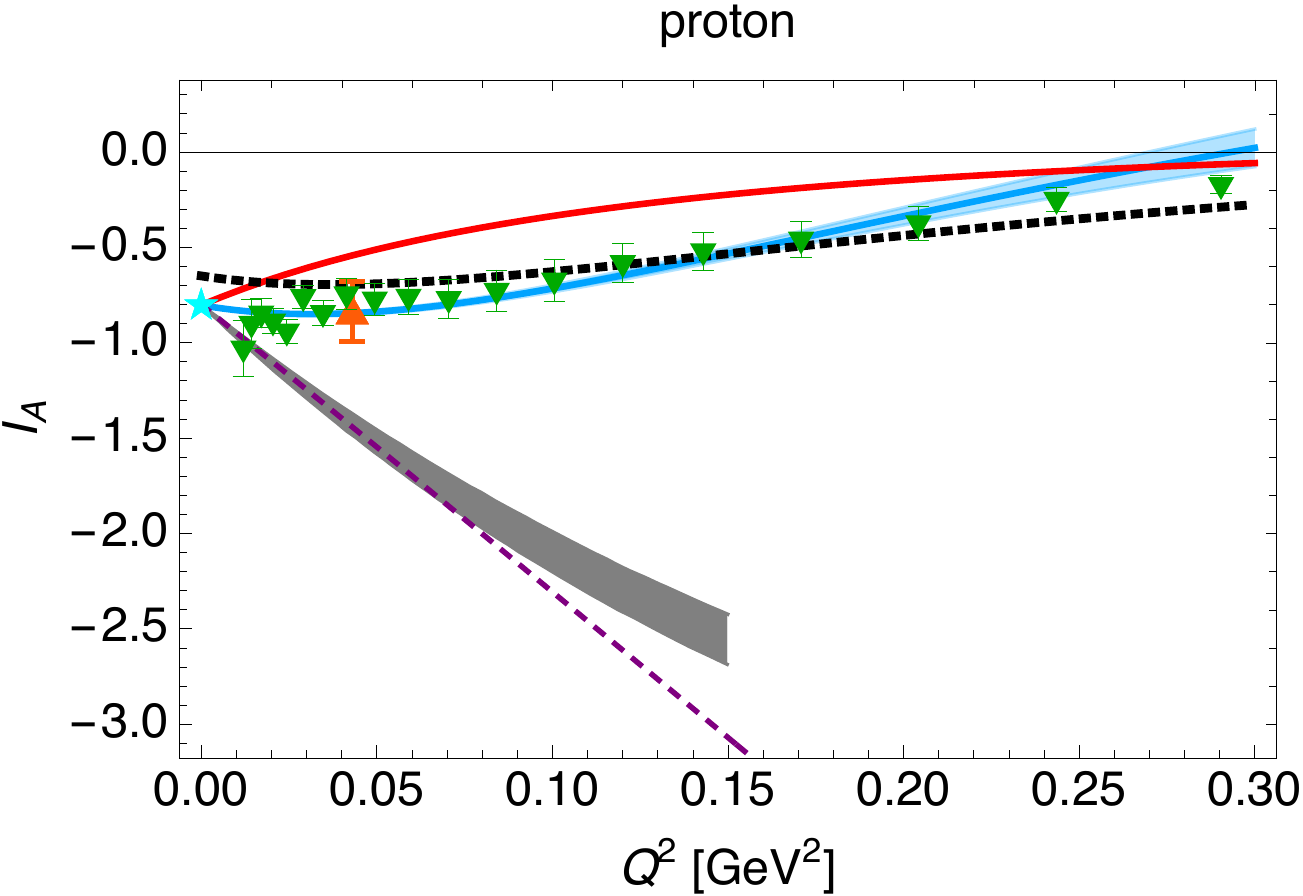}\\
\caption{Generalized GDH integrals $I_1 (Q^2)$ and $I_A (Q^2)$ of the proton. Legend is the same as in Fig.~\ref{fig:protonspin}.
Data-driven evaluation: The cyan star at $Q^2=0$ is the GDH value: 
$-\nicefrac14 \kappa_p^2$,
using  $\kappa_p \approx 1.793$. 
 \label{fig:CLASdata}   }
\end{center}
\end{figure}

 In this regard, it is interesting to consider the following integrals, 
\begin{subequations}
\bea
I_1(Q^2)&=&\frac{2M_N^2}{Q^2}\int_0^{x_0}\dd x\, g_1(x,Q^2),\eqlab{I1def}\\
\eqlab{genGDHnonpole}
I_A(Q^2)&=&\frac{2M_N^2}{Q^2}\int_0^{x_0}\dd x\, \left[g_1(x,Q^2)-\frac{4M_N^2 x^2}{Q^2}\,g_2(x,Q^2)\right].
\eea
\end{subequations}
These are generalizations of the GDH sum rule. The latter holds at the real-photon point: $I_1(0)= I_A(0) = - \nicefrac14\, \kappa_N$, with $\kappa_N$ being the nucleon anomalous magnetic moment. The high-energy (low-$x$) contributions in these integrals are not suppressed by the factor $x^2$, and hence, the differences with MAID seen in polarizabilities should be amplified here. Certainly, MAID is missing here the higher-energy contributions needed to saturate the GDH sum rule, as is clearly seen at $Q^2=0$ in \Figref{CLASdata}.
Yet, these missing contributions are not very large; of the order of tens of percent. Deviations from MAID by factors 2 or 3, seen by the Bochum calculations 
in \Figref{CLASdata}, are hardly explainable. Even less so are the large deviations in polarizabilities in \Figref{protonspin}, especially in $\de_{LT}$ of the proton.

The same sentiment applies to the new data-driven evaluation of neutron polarizabilities by 
the E97-110 Collaboration \cite{E97-110:2021mxm}, shown in \Figref{protonspin} by pink crosses. It has anomalously large deviations from MAID in the low-$Q$ region.   
The new CLAS results for the proton \cite{CLAS:2021apd} (green triangles) are, on the other hand, more reasonable.
Hence, the anomalous results for the neutron is likely due to complications arising in extracting the neutron properties from ''neutron targets'', e.g., 
the deuteron or helium-3. 

New data for the proton from the g2p Collaboration, presented at this workshop, have sofar  been 
shown in a preprint \cite{JeffersonLabHallAg2p:2022qap};  see Figs.~2 and 4  therein, for a comparison of theory predictions and data for the proton 
$\delta_{LT}(Q^2)$ and the inelastic moment $\bar d_2(Q^2)$. The latter is defined as:
\begin{align}\label{Eq:d2WW}
 \bar d_2(Q^2) =  \int^{x_0}_0 \!\! \dd x\ x^2 \,[3 g_2(x,Q^2) + 2 g_1(x,Q^2)], 
\end{align}
and relates to the inelastic part of the twist-3 part of the spin structure function $g_2(x,Q^2)$ \cite{Jaffe:1989xx,Shuryak:1981pi}.
Note that this quantity can be derived from the aforementioned quantities, e.g., 
\beq 
\bar d_2(Q^2)  = \frac{Q^6}{8\al M_N^2} \delta_{LT}(Q^2) + \frac{Q^4}{8\al M_N^4} \big[ I_1 (Q^2) - I_A (Q^2) \big],
\eeq
and hence does not add anything new to this discussion.

\section{Polarizability Contribution to the Hyperfine Splitting in (Muonic-)Hydrogen }\seclab{hfsChPTSEC}

The hfs of the $nS$-level  is proportional to the leading order-$\alpha^4$ Fermi energy:
\beq
E_\mathrm{F}=\frac{8\,\al^4 m_r^3}{3mM}(1+\kappa_p)\eqlab{FermiE},
\eeq
where $M$ is the proton mass, $\kappa_p$ is the anomalous magnetic moment of the proton, $m$ is the electron or muon mass in the case of H or $\mu$H, respectively, and $m_r=mM/(m+M)$ is the reduced mass.
The nuclear finite-size effects start contributing at the order $\alpha^5$, through the forward \2PE exchange, \Figref{TPE},
which is conventionally split into the elastic (Zemach-radius and recoil) and inelastic (polarizability) contributions \cite{Carlson:2008ke}:
\beq
E^\mathrm{\mTPE}_\mathrm{hfs}(nS)=\frac{E_\mathrm{F}}{n^3}\left(\Delta_\mathrm{Z}+\Delta_\mathrm{recoil}+\Delta_\mathrm{pol}\right). \eqlab{FS_hfs}
\eeq
The finite-size contributions are obtained from the proton electromagnetic form factors, see, e.g., Ref.~\cite{Antognini:2022xqf} for a recent update of the recoil contribution.

The polarizability contribution to hfs is expressed in terms of the spin structure functions:
\begin{subequations}
\eqlab{POL}
\beq
\Delta_\mathrm{pol}=\Delta_1+\Delta_2=\frac{\al m}{2\pi (1+\kappa_p) M}\left[\delta_1+\delta_2\right],
\eeq
where $\Delta_1$ and $\Delta_2$ are related to the spin-dependent structure functions $g_1$ and $g_2$, respectively:
\begin{align}
\delta_1
&=2\int_0^\infty\dd Q\left(k_0(Q^2)\left[4I_1(Q^2)+F_2^2(Q^2)\right]+\int_0^{x_0}\dd x\,k_1(x,Q^2)\, g_1(x,Q^2)\right),\eqlab{Delta1b}\\
\delta_2&=\int_0^\infty\dd Q\int_0^{x_0}\dd x\, k_2(x,Q^2)\,g_2(x,Q^2),\eqlab{Delta2}
\end{align}
\end{subequations}
and the kinematic functions $k_i$ can be found in, e.g., \cite{Antognini:2022xoo}.
In \Eqref{Delta1b}, we isolated the polarizability part, $I_1^\mathrm{(pol)} = I_1(Q^2)+ \nicefrac14 F_2^2(Q^2) $, of the first moment of the $g_1$ spin structure function. This is not required to achieve convergence in the dispersive description of the $S_1$ amplitude. However,  it is important because $4I_1(Q^2)$ and $F_2(Q^2)$  cancel 
exactly at $Q^2=0$, thanks to the GDH sum rule. A large cancellation persists also at finite momentum transfer. With the Pauli form factor $F_2$ and $g_1$ structure function parametrizations of Refs.\ \cite{Kelly:2004hm} and \cite{Simula:2001iy}, we find contributions to $\Delta_\mathrm{pol}$ of $1089$ ppm and $-855$ ppm, respectively. Their cancellation into $234$ ppm is a considerable source of uncertainty in the data-driven evaluation. Evaluations of the full polarizability contribution, based on similar parametrizations, yield $\Delta_\mathrm{pol}$ more than a factor $2$ smaller than the individual $I_1(Q^2)$ and $F_2(Q^2)$ contributions. A low-$Q$ expansion of $\Delta_1$ is conventionally used to interpolate between the real-photon limit, described by the static values of the forward spin polarizability $\gamma_0$ and the anomalous magnetic moment $\kappa$, and the onset of data for the $g_1$ structure function, see, e.g., Ref.~\cite{Carlson:2008ke}.

\begin{figure}[tbh]
\centering
       \includegraphics[width=13cm]{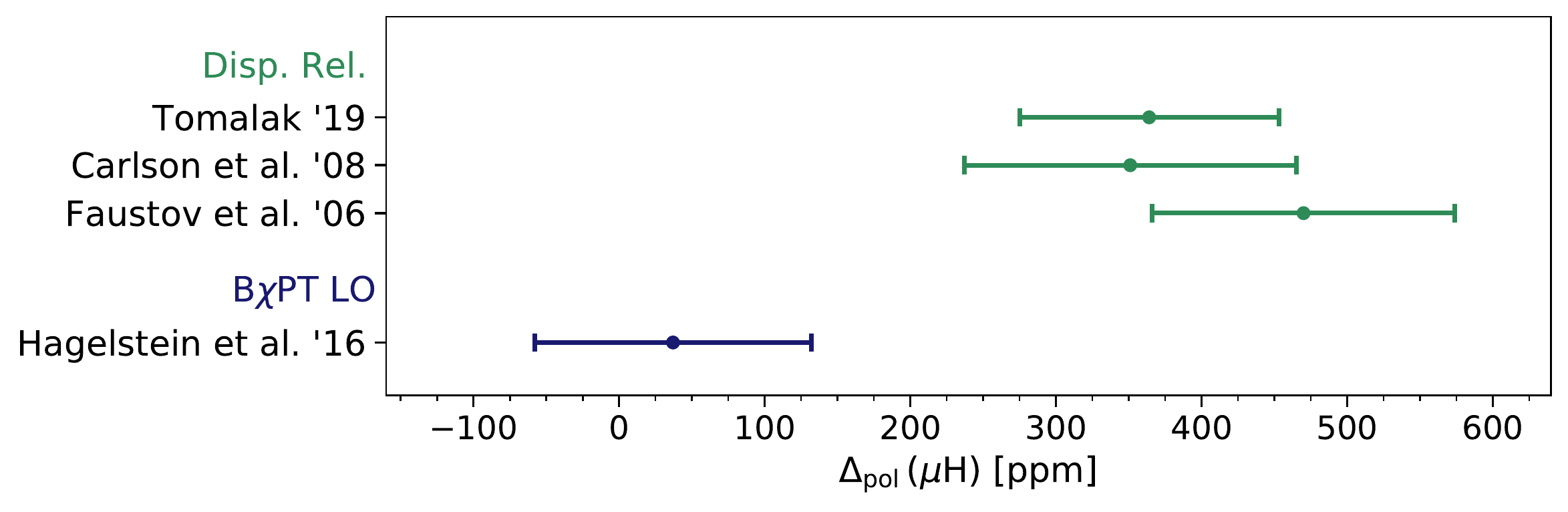}
\caption{Comparison of predictions for the polarizability contribution to the hfs $\mu$H \cite{Carlson:2011af,Faustov:2006ve,Tomalak:2018uhr}. \label{fig:hfspol}}
\end{figure}

Presently, there is a discrepancy between the data-driven dispersive evaluations and the B$\chi$PT predictions of the polarizability contributions to the hfs in $\mu$H, see Fig.~\ref{fig:hfspol}. Similar discrepancy exists in H.
The  B$\chi$PT prediction is considerably smaller and is, in fact, comparable with $0$ \cite{Hagelstein:2015lph}. This can be understood from a low-energy expansion of the spin-dependent VVCS amplitudes in the heavy-baryon (HB) limit~\cite{Pineda:2002as}, where the leading HB term, $O(1/m_\pi^2)$, cancels out in $\ol S_1(0,Q^2)$. Therefore, the chiral loops in the hfs are essentially vanishing, where the small number is just a remnant of higher orders in the HB expansion.  

\begin{figure}[thb]
\centering
 \includegraphics[width=0.95\textwidth]{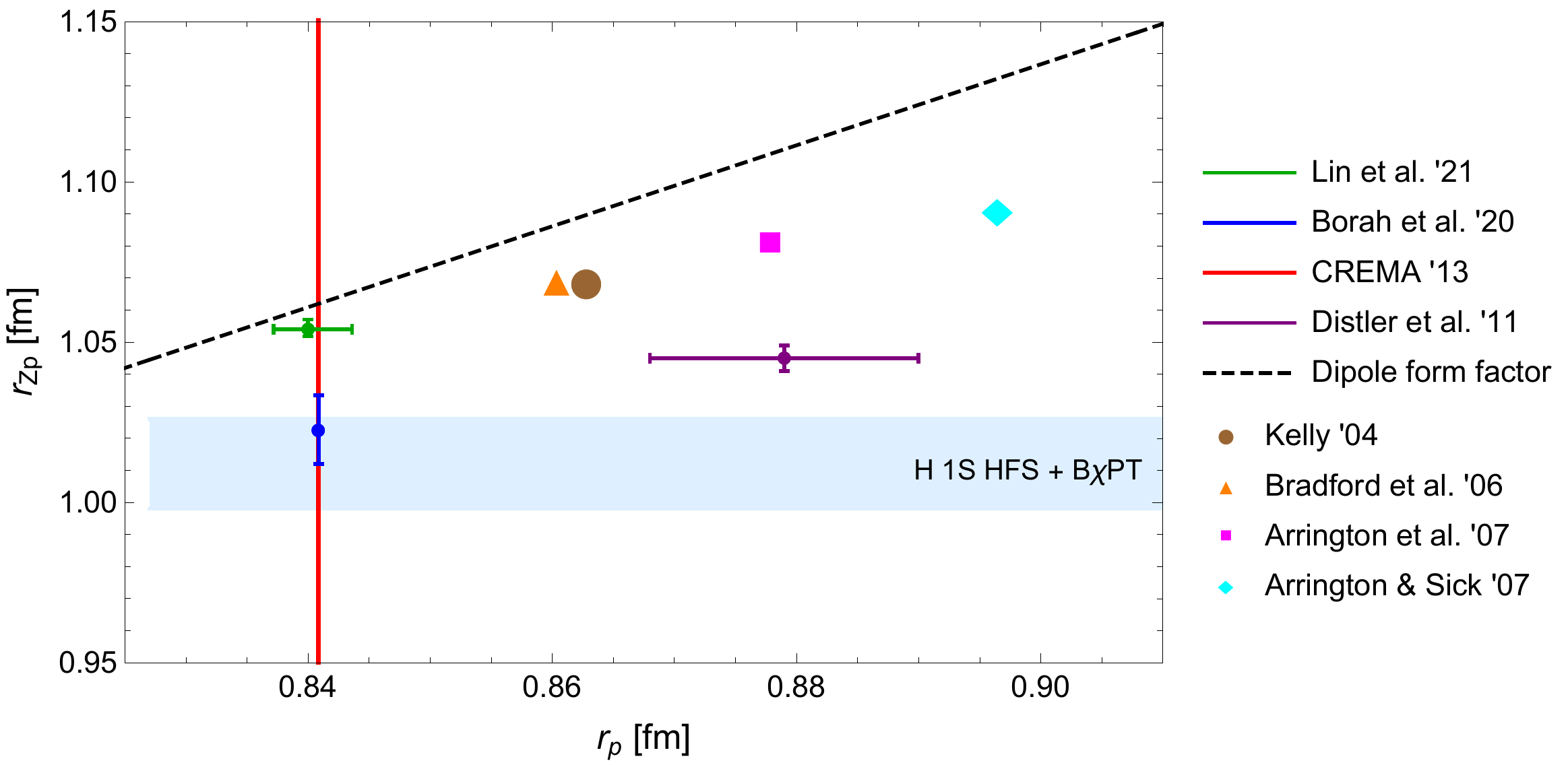}
\caption{Correlation between the Zemach and charge radius of the proton. The points are from: Lin et al.~\cite{Lin:2021xrc}, Borah et al.~\cite{Borah:2020gte}, CREMA \cite{Antognini:1900ns}, Distler et al.~\cite{Distler:2010zq}, Kelly \cite{Kelly:2004hm}, Bradford et al.~\cite{Bradford:2006yz}, Arrington et al.~\cite{Arrington:2007ux}, and Arrington \& Sick \cite{Arrington:2006hm}. Figure is taken from \cite{Antognini:2022xoo}.
        \label{fig:RadiiCorrelation}}
\end{figure}

Since the total \2PE contribution is well constrained by the precise measurement of hydrogen hfs, the smaller polarizability effect in B$\chi$PT implies a smaller Zemach radius, cf.\ the blue band in \Figref{RadiiCorrelation} and Ref.~\cite{Antognini:2022xoo} for more details.

\section{Conclusion} \seclab{Sum}
We have discussed two  discrepancies 
in the current description of the nucleon spin structure at low energies.
\begin{itemize}
    \item The discrepancy between the two 
    B$\chi$PT calculations (Mainz vs.\ Bochum) of the nucleon spin polarizabilities $\de_{LT}$ and $\gamma_0$, and generalized GDH integrals.
    \item The discrepancy between B$\chi$PT and data-driven evaluations of the proton polarizability contribution to hfs, see Fig.~\ref{fig:hfspol}.
\end{itemize}
To resolve these discrepancies one does not necessarily need new experimental data, there are plenty of data against which these calculations have not been tested yet. 

For example, there is a wealth of empirical information on real Compton scattering of the proton. In the forward kinematics everything is
known \cite{Gryniuk:2015eza,Gryniuk:2016gnm}, the value of $\ga_0(0)$, seen in \Figref{protonspin}, is only one of the data points.  The off-forward Compton scattering has also been well-measured
and can serve as a test of these calculations, see e.g., \cite{Lensky:2015awa}.

The data-driven evaluations  of the polarizability 
contribution to the hyperfine splitting ought to set some benchmarks as well. They could, for example, compute the proton spin polarizabilities and the aforementioned GDH integrals using the same ingredients. 

 The forthcoming re-analysis of $g_2(x,Q^2)$,  and its contribution to hfs,
 by the g2p Collaboration is promising to improve the situation \cite{Zielinski:2017gwp,JeffersonLabHallAg2p:2022qap}.  
 They, first of all, obtain new data for this otherwise scarcely-known spin structure function.
And they will provide the results for the spin
polarizabities and their hfs contribution in the same evaluation.
\acknowledgments

This work is supported by the Swiss National Science Foundation (SNSF) through the Ambizione Grant PZ00P2\_193383, the Deutsche Forschungsgemeinschaft (DFG) 
through the Emmy Noether Programme under the grant 449369623 and through the project 204404729-SFB1044. 


\begin{small}
\bibliographystyle{JHEP}
\bibliography{lowQ}
\end{small}

\end{document}